%% file: main.tex
\documentclass[11pt]{article}

\usepackage[english]{babel}
\usepackage[utf8x]{inputenc}
\usepackage[T1]{fontenc}
\usepackage{authblk}
\usepackage{bbm}

\sffamily
\usepackage[a4paper,top=2cm,bottom=2cm,left=2cm,right=2cm,marginparwidth=1.75cm]{geometry}

\usepackage[tbtags]{amsmath}
\usepackage{amsfonts}
\usepackage{dsfont}
\usepackage{soul}
\usepackage[colorlinks=true, allcolors=blue]{hyperref}
\usepackage{natbib}
\setcitestyle{authoryear,open={(},close={)}}

\author[1]{Anirudh Tomer}
\author[1]{Jorne Biccler}
\affil[1]{P95 Epidemiology \& Pharmacovigilance, Leuven, Belgium}

\title{Temporal Properties of Vaccine Effectiveness Measures in Presence of Multiple Pathogen Variants and Multiple Vaccines}
\date{\vspace{-7ex}}

\begin{document}
\maketitle

\input{contents/0_abstract}
\input{contents/1_introduction}

\input{contents/2_ve_estimands}
\input{contents/3_estimand_components}
\input{contents/4_ve_time_dependency}
\input{contents/5_test_negative_design}
\input{contents/6_sample_size}
\input{contents/7_discussion}
\input{contents/8_acknowledgments}

\section*{Funding}
The DRIVE project has received funding from the Innovative Medicines Initiative (IMI) 2 Joint Undertaking under grant agreement No 777363. This joint undertaking receives support from the European Union´s Horizon 2020 research and innovation and European Federation of Pharmaceutical Industries and Associations (EFPIA). The IMI is a joint initiative (public-private partnership) of the European Commission and EFPIA to improve the competitive situation of the European Union in the field of pharmaceutical research. The IMI provided support in the form of salaries for Anirudh Tomer and Jorne Biccler but did not have any additional role in the study design, data collection and analysis, decision to publish, or preparation of the manuscript.

\appendix
\section*{Appendix}
\input{contents/appendix}

\bibliographystyle{apalike}
\bibliography{main}

\end{document}

%% file: contents/0_abstract.tex
\begin{abstract}
Vaccine effectiveness (VE) is typically defined as incidence rate ratio, cumulative-risk ratio, or odds ratio. Under certain conditions, the VE based on incidence rate ratio is known to be time-invariant over the study period for leaky action vaccines. In contrast, the VE based on cumulative-risk ratio is time-invariant for all-or-none action vaccines. Consequently, these two VE measures are recommended as appropriate measures of VE for leaky and all-or-none vaccines, respectively. However, in diseases with multiple pathogen variants (e.g., influenza, COVID-19, dengue, human immunodeficiency virus) and multiple vaccines, investigators may also be interested in variant-specific VE of a vaccine, the relative VE of a vaccine against two variants, or the relative VE of different vaccines against a given variant. In this multi-variant and multi-vaccine scenario, the temporal properties of the aforementioned VE measures have not been studied entirely yet. Furthermore, no general-purpose sample size calculator is available for either studies that intend to estimate variant-specific VE or relative VE. Our motivation to address these two challenges comes from the European Union's Innovative Medicines Initiatives' DRIVE project and the current COVID-19 pandemic. 

As a solution, we first define variant-specific and relative VE measures while accounting for multiple competing pathogen variants. We then propose a generic mode of action for all-or-none vaccines in our multi-variant setting. Subsequently, we evaluate the conditions and the extent to which various VE measures can be time-varying. We show that every VE measure is time-varying for all-or-none action vaccines in a multi-variant and multi-vaccine scenario. On the other hand, we show that all measures other than those based on incidence rate ratios are time-varying for leaky action vaccines. We discuss the practical implications of these results on VE studies in the context of the commonly used cohort, cumulative case-control, and test-negative study designs. Lastly, for the multi-variant and multi-vaccine scenario, we implement sample size calculations for both variant-specific and relative VE in an \textbf{R} package and an online calculator.
\end{abstract}

%% file: contents/1_introduction.tex
\section{Introduction}
\label{sec:intro}
A key quantity of interest in vaccine studies is vaccine effectiveness for susceptibility to infection (\textit{VE}). It pertains to the protective effects of vaccination against a pathogen. When a pathogen is genotypically and/or phenotypically diverse, differential VE against its \textit{variants} is of interest too \citep{dengue_vaccine, lopez2021effectiveness}. An augmentation of this situation occurs when there is more than one vaccine for a pathogen and vaccines are compared against each other, for example, as in the COVID-19 pandemic \citep{gupta2021will}. In such scenarios, both variant-specific VE and relative VE of vaccines may need to be calculated. The VE is typically estimated as the percentage decrease in relative risk of infection among the vaccinated subjects over the unvaccinated subjects \citep{halloran2010design}. This relative-risk in most real-world situations is either defined as incidence rate (or incidence density) ratio ($irr$), hazard rate ratio ($hr$), or cumulative-risk (or cumulative-incidence) ratio ($crr$). Using these, different \textit{measures} of VE can be defined as $\mbox{VE}^{irr} = 1 - irr$, $\mbox{VE}^{hr} = 1 - hr$, and $\mbox{VE}^{crr} = 1 - crr$, each having a different interpretation. 

The choice of an appropriate VE measure may depend upon the goal of the study and also on the action mechanism (leaky or all-or-none) of the vaccine \citep[page 132]{halloran2010design}. In this regard, \citet{smith1984assessment} have proposed using $\mbox{VE}^{irr}$ over $\mbox{VE}^{crr}$ in leaky action vaccines because, in leaky vaccines $\mbox{VE}^{irr}$ remains constant over the study period whereas $\mbox{VE}^{crr}$ does not. The opposite of this phenomenon happens in all-or-none action vaccines; therefore, $\mbox{VE}^{crr}$ is recommended as the VE measure there. A vaccine's action mechanism and the temporal properties of a VE measure are also linked with study designs and VE estimation. For example, in case-control \citep{breslow1996statistics} and test-negative design (TND) studies, either $\mbox{VE}^{irr}$ or $\mbox{VE}^{crr}$ can be estimated by utilizing $1 - \hat{or}$, where $\hat{or}$ is the exposure odds ratio in the sample. Here, whether the temporal properties of $\mbox{VE}^{irr}$ or those of $\mbox{VE}^{crr}$ will apply depends on the approach for sampling controls \citep{dean2019re, lewnard2018measurement}. We next expand upon how the various VE measures and their temporal properties are interesting from both the vaccine manufacturers and public health perspectives and our related motivations and goal of this work.

The motivation for this work comes from two sources. First, from the requirements of the Development of Robust and Innovative Vaccine Effectiveness (DRIVE, \url{https://www.drive-eu.org/}) project of Innovative Medicines Initiative - IMI \citep{IMIproject} which provided funding for this work. The DRIVE project aims to set up a platform, bringing together all stakeholders to study the brand and variant-specific influenza vaccine effectiveness in the European Union. Within DRIVE, both cohort and TND studies are used. For these designs, in the context of DRIVE's multi-variant and multi-vaccine scenario, two challenges are as follows. First, are the temporal properties of variant-specific VE measures comparable to the temporal properties of VE in a single variant scenario \citep{smith1984assessment}? This may affect conclusions regarding the effectiveness of a vaccine. The time-dependency of a VE measure is not a drawback per se in other situations. For example, $\mbox{VE}^{crr}$ is time-dependent in leaky vaccines, but it can still be useful for assessing the population level benefit of a vaccine rollout. This brings us to the second challenge: no sample size calculator that handles multiple variants and vaccines is available for study planners. While the DRIVE project focuses on influenza, the multi-variant situation is also present in other pathogens such as dengue~\citep{dengue_vaccine} and human-immunodeficiency virus~\citep{gilbert1998statistical}. Our second motivation comes from the current COVID-19 pandemic~\citep{world2020coronavirus}. Recently, several studies that estimate variant-specific VE for COVID-19 vaccines~\citep{zimmer_corum_wee_2020} have been published~\citep{shinde2021efficacy,lopez2021effectiveness, abu2021effectiveness,madhi2021efficacy}. Since these studies focus on the change of VE over time, it is crucial to know if the variation of VE is, in fact, the biological effect of the vaccine or whether this could be an artifact due to the use of a time-dependent VE measure. Hence, these studies show that knowledge on temporal properties of $\mbox{VE}^{irr}, \mbox{VE}^{crr}, \mbox{VE}^{or}$ is of importance, and substantiate the need for a sample size calculator.

The aim of this work is two-fold. First, to extend upon the work of \citet{smith1984assessment} to study the temporal properties of VE measures in the presence of multiple variants and vaccines, for both leaky and all-or-none action vaccines, in cohort and TND study designs. For this, we not only focus on $\mbox{VE}^{irr}$ and $\mbox{VE}^{crr}$ but also check if $\mbox{VE}^{or} = 1-or$ is time-invariant. The $\mbox{VE}^{or}$ is rarely the target measure; however, we study it because when the rare events assumption is not met in case-control studies based on the inclusive sampling scheme, the sample exposure odds ratio based VE estimator $1 - \hat{or}$ does not estimate $\mbox{VE}^{crr}$. Instead, it estimates $\mbox{VE}^{or}$, which can be seen as a measure of VE based on the odds ratio. We also compare the use of $1 - irr$, $1 - crr$, and $1 - or$ as measures of relative VE of vaccines. Our second goal is to implement calculations for obtaining power and precision \citep{kelley2003obtaining} for VE in the multi-variant and multi-vaccine scenario, in an \textbf{R} package and an online calculator.

Previously, for cohort studies, \citet{gilbert2000comparison} studied the temporal properties and proposed estimators for VE of a single leaky vaccine against multiple variants. However, to our knowledge, no such work exists for all-or-none action vaccines. In TND studies, \citet{lewnard2018measurement} and \citet{dean2019re} have studied the temporal properties of the estimator $1 - \hat{or}$, with two different control sampling techniques, namely inclusive sampling and incidence density sampling. Both authors have corroborated the findings of \citet{smith1984assessment}, but only for a single vaccine and single pathogen variant. However, within TND studies, there is still scope for discussing the scenario of multiple competing variants and the relative VE of vaccines. Regarding our sample size calculation goal for VE, currently available tools such as the \textbf{R} package \textbf{epiR} \citep{epiR} and general-purpose calculators such as PASS (\url{https://www.ncss.com/software/pass/}) have two drawbacks. Specifically, they only handle the single variant and single vaccine scenario, and for sample size calculations aiming at obtaining a certain precision \citep{kelley2003obtaining} for VE, the calculators ignore the randomness of the data generating process.

The rest of this article is as follows. In Section~\ref{sec:ve_measures}, we define $\mbox{VE}^{irr}, \mbox{VE}^{crr}, \mbox{VE}^{or}$ employing a generic scenario of $M$ vaccines and $I$ variants in a cohort study. In Section~\ref{sec:ve_components}, we define the probability components of these measures for subjects vaccinated with a leaky vaccine, an all-or-none action vaccine, or placebo. In Section~\ref{sec:ve_time_dependency}, we study the temporal properties of the VE measures for leaky and all-or-none action vaccines, and in Section~\ref{sec:tnd}, we discuss them further for TND studies. In Section~\ref{sec:samp_size}, we present a sample size calculator for the multi-variant and multi-vaccine scenario. Lastly, Section~\ref{sec:discussion} entails the discussion of this work.

%% file: contents/2_ve_estimands.tex
\section{Measures of VE}
\label{sec:ve_measures}
Consider a set $\mathcal{M} = \{1, \ldots, M\}$ of vaccines of interest whose VE we intend to calculate against a set $\mathcal{I} = \{1, \ldots, I\}$ of variants of a pathogen circulating in the source population. The VE measures described in this manuscript focus on describing the vaccine effect at the end of a follow up-period of length $t$ in a population for which no immunity was present at the beginning of the study. In what follows, we will rely on a simplification of the assumptions mentioned by \citet{gilbert1998statistical} and \citet{greenwood1915}. More specifically, we assume that the vaccine assignment was randomized, the different strains can be considered to be competing events, and for all subjects the incidence rate of infection from any strain remains constant throughout the follow-up period.

Suppose that among unvaccinated subjects, the incidence rate of infection due to a variant $i \in \mathcal{I}$ is given by $\lambda_i > 0$. We assume $\lambda_i$ to be constant over the study period for ease of exposition. Subsequently, let the overall rate of infection due to all variants be $\Lambda=\sum_{i\in \mathcal{I}} \lambda_i$. At the end of the study period, let $C=i$ indicate getting infected (case) by variant $i$ and $C=0$ indicate remaining uninfected (control), $V=m$ denotes being vaccinated by vaccine $m \in \mathcal{M}$ and $V=0$ denotes receiving placebo. Let $Y$ represent the person-time contribution of a subject in the study. It is defined as the minimum of the time to infection and the study end time $t$. Then, in a cohort study, the data available at the end of the study is obtained with the probabilities shown in Table~\ref{tab:available_data}, and the expected person-time contribution shown in the last row of Table~\ref{tab:available_data}.

\begin{table}[]
\centering
\begin{tabular}{| c | c | c | c | c |} 
 \hline
  &  \textbf{Placebo} ($V=0$) & \textbf{Vaccine~$1$} ($V=1$) & $\ldots$ & \textbf{Vaccine~$M$} ($V=M$) \\ 
 \hline
 \textbf{Uninfected} ($C=0$) & $\mathds{P}(C=0 \mid V=0)$ & $\mathds{P}(C=0 \mid V=1)$ & $\ldots$ & $\mathds{P}(C=0 \mid V=M)$ \\ 
 \textbf{Variant~$1$} ($C=1$) & $\mathds{P}(C=1 \mid V=0)$ & $\mathds{P}(C=1 \mid V=1)$ & $\ldots$ & $\mathds{P}(C=1 \mid V=M)$\\ 
  $\vdots$ & $\vdots$ & $\vdots$ & $\ddots$ & $\vdots$ \\
  \textbf{Variant~$I$} ($C=I$) & $\mathds{P}(C=I \mid V=0)$ & $\mathds{P}(C=I \mid V=1)$ & $\ldots$ & $\mathds{P}(C=I \mid V=M)$\\
 \hline
 \textbf{Person-time $Y$ contributed} & $E(Y \mid V=0)$ & $E(Y \mid V=1)$ & $\ldots$ & $E(Y \mid V=M)$\\
 \hline
\end{tabular}
\caption{Cell probabilities of the data available from a cohort study with multiple variants and multiple vaccines. Here, $C=i$ indicates getting infected (case) by variant $i \in \mathcal{I}$ and $C=0$ indicates remaining uninfected (control), $V=m$ denotes being vaccinated by vaccine $m \in \mathcal{M}$ and $V=0$ denotes receiving placebo. Let $Y$ represent the person-time contribution of a subject in the study. It is defined as the minimum of the time to infection and the study end time $t$. The probabilities $\mathds{P}(C=i \mid V=\cdot)$ and $\mathds{P}(C=0 \mid V=\cdot)$ are the cumulative-risks of getting infected and not getting infected, respectively, by the variant $i$ in subjects vaccinated with the vaccine $m$ (or placebo). The ${E(Y \mid V=\cdot)}$ denotes the expected person-time a subject vaccinated with the vaccine $m$ (or placebo) will spend in the study.}
\label{tab:available_data}
\end{table}

\subsection{Variant-specific VE of Vaccine $m$ Against Variant $i$}
Our first aim is to define the VE of vaccine $m$ against the variant $i$ compared to the placebo group. We call this VE the variant-specific VE of the vaccine and utilize three different measures for it. These are namely, $\mbox{VE}_{i,m}^{irr}$ based on incidence rate ratio, $\mbox{VE}_{i,m}^{crr}$ based on ratio of cumulative-risks, and $\mbox{VE}_{i,m}^{or}$ based on exposure odds ratio. Using Table~\ref{tab:available_data}, these three are defined as:
\begin{align}
\label{eq:ve_absolute_def}
\begin{split}
    \mbox{VE}_{i,m}^{irr} &= 1 - \frac{\mathds{P}(C=i \mid V=m)}{E(Y \mid V=m)} \div \frac{\mathds{P}(C=i \mid V=0)}{E(Y \mid V=0)},\\
    \mbox{VE}_{i,m}^{crr} &= 1 - \frac{\mathds{P}(C=i \mid V=m)}{\mathds{P}(C=i \mid V=0)},\\
    \mbox{VE}_{i,m}^{or} &= 1 - \frac{\mathds{P}(V=m \mid C=i)}{\mathds{P}(V=0 \mid C=i)} \div \frac{\mathds{P}(V=m \mid C=0)}{\mathds{P}(V=0 \mid C=0)}\\
    &= 1 - \frac{\mathds{P}(C=i \mid V=m)}{\mathds{P}(C=i \mid V=0)} \div \frac{\mathds{P}(C=0 \mid V=m)}{\mathds{P}(C=0\mid V=0)},
\end{split}
\end{align}
where the probabilities $\mathds{P}(C=i \mid V=\cdot)$ and $\mathds{P}(C=0 \mid V=\cdot)$ are the cumulative-risks of getting infected and not getting infected, respectively, by the variant $i$ in subjects vaccinated with the vaccine $m$ (or placebo). The ${E(Y \mid V=\cdot)}$ denotes the expected person-time a subject vaccinated with the vaccine $m$ (or placebo) will spend in the study. In addition, $\mathds{P}(C=i \mid V=\cdot) / E(Y \mid V=\cdot)$ denotes the incidence rate of infection with variant $i$ among subjects with the given vaccination status. It is important to note that for $\mbox{VE}_{i,m}^{or}$ in (\ref{eq:ve_absolute_def}), we expressed the odds ratio $or$ retrospectively by conditioning on the infection status $C$. Specifically, we first used the probablities $\mathds{P}(V=m \mid C=\cdot)$ and later converted them using the Bayes' theorem to $\mathds{P}(C=i \mid V=\cdot)$. We did so because $or$ is typically employed in case-control studies wherein data is collected retrospectively on vaccination status $V$, for a known infection status $C$.

\subsection{Relative VE of Vaccine $m$ Against Two Variants $i$ and $j$}
Our next aim is to compare the relative VE of a vaccine $m$ against two variants, $i$ and $j$. To this end, three measures of the relative VE, namely $\mbox{VE}_{ij,m}^{irr}$ based on incidence rate ratio, $\mbox{VE}_{ij,m}^{crr}$ based on ratio of cumulative-risks, and $\mbox{VE}_{ij,m}^{or}$ based on exposure odds ratio are defined as:
\begin{align}
\label{eq:ve_relative_given_vaccine_def}
\begin{split}
    \mbox{VE}_{ij,m}^{irr} &= 1 - \frac{1 - \mbox{VE}_{i,m}^{irr}}{1 - \mbox{VE}_{j,m}^{irr}}
    = 1 - \frac{\mathds{P}(C=i \mid V=m)}{\mathds{P}(C=j \mid V=m)} \div \frac{\mathds{P}(C=i \mid V=0)}{\mathds{P}(C=j \mid V=0)},\\
    \mbox{VE}_{ij,m}^{crr} &= 1 - \frac{1 - \mbox{VE}_{i,m}^{crr}}{1 - \mbox{VE}_{j,m}^{crr}}
    = 1 - \frac{\mathds{P}(C=i \mid V=m)}{\mathds{P}(C=j \mid V=m)} \div \frac{\mathds{P}(C=i \mid V=0)}{\mathds{P}(C=j \mid V=0)},\\
    \mbox{VE}_{ij,m}^{or} &= 1 - \frac{1 - \mbox{VE}_{i,m}^{or}}{1 - \mbox{VE}_{j,m}^{or}}
    = 1 - \frac{\mathds{P}(V=m \mid C=i)}{\mathds{P}(V=0 \mid C=i)} \div \frac{\mathds{P}(V=m \mid C=j)}{\mathds{P}(V=0 \mid C=j)}\\
    &= 1 - \frac{\mathds{P}(C=i \mid V=m)}{\mathds{P}(C=j \mid V=m)} \div \frac{\mathds{P}(C=i \mid V=0)}{\mathds{P}(C=j \mid V=0)}.
\end{split}
\end{align}
All three definitions lead to the same result, that is, a ratio of the odds of being a case of variant $i$ versus the variant $j$ in vaccinated subjects and the same odds in placebo subjects. 

\subsection{Relative VE of Two Vaccines $m$ and $n$ Against a Variant $i$}
Last, we define the relative VE of two vaccines $m$ and $n$ against the same variant $i$. For this purpose, three measures of the relative VE, namely $\mbox{VE}_{i,mn}^{irr}$ based on incidence rate ratio, $\mbox{VE}_{i,mn}^{crr}$ based on ratio of cumulative-risks, and $\mbox{VE}_{i,mn}^{or}$ based on exposure odds ratio are defined as:
\begin{align}
\label{eq:ve_relative_given_variant_def}
\begin{split}
    \mbox{VE}_{i,mn}^{irr} &= 1 - \frac{1 - \mbox{VE}_{i,m}^{irr}}{1 - \mbox{VE}_{i,n}^{irr}}
    = 1 - \frac{\mathds{P}(C=i \mid V=m)}{E(Y \mid V=m)} \div \frac{\mathds{P}(C=i \mid V=n)}{E(Y \mid V=n)},\\
    \mbox{VE}_{i,mn}^{crr} &= 1 - \frac{1 - \mbox{VE}_{i,m}^{crr}}{1 - \mbox{VE}_{i,n}^{crr}}
    = 1 - \frac{\mathds{P}(C=i \mid V=m)}{\mathds{P}(C=i \mid V=n)},\\
    \mbox{VE}_{i,mn}^{or} &= 1 - \frac{1 - \mbox{VE}_{i,m}^{or}}{1 - \mbox{VE}_{i,n}^{or}}
     = 1 - \frac{\mathds{P}(V=m \mid C=i)}{\mathds{P}(V=n \mid C=i)} \div \frac{\mathds{P}(V=m \mid C=0)}{\mathds{P}(V=n \mid C=0)}\\
    &= 1 - \frac{\mathds{P}(C=i \mid V=m)}{\mathds{P}(C=i \mid V=n)} \div \frac{\mathds{P}(C=0 \mid V=m)}{\mathds{P}(C=0\mid V=n)}.
\end{split}
\end{align}

We next define the component probabilities of the measures in (\ref{eq:ve_absolute_def}), (\ref{eq:ve_relative_given_vaccine_def}), and (\ref{eq:ve_relative_given_variant_def}) among the placebo and vaccinated subjects, for both leaky and all-or-none action vaccines.

%% file: contents/3_estimand_components.tex
\section{VE Components: Cumulative-Risk, Odds, and Person-time}
\label{sec:ve_components}
\subsection{Placebo Subjects}
Since we assumed that infection can happen with only one variant over the study period, infections from all variants are competing events \citep{putter2007tutorial}. Under the competing events framework, among placebo subjects ($V=0$) the cumulative-risk of being a control ($C=0$), the cumulative-risk of being a case of the variant $i$ ($C=i$), and the expected person-time contribution of a subject $E(Y \mid V=0)$ over the study period $[0, t]$ is given by:
\begin{align}
\label{eq:placebo_components}
\begin{split}
\mathds{P}(C=0 \mid V=0) &= \exp(-\Lambda t),\\
\mathds{P}(C=i \mid V=0) &= \frac{\lambda_i}{\Lambda} \big\{1 - \exp(-\Lambda t)\big\},\\
E(Y \mid V=0) &= \int_0^t \exp(-\Lambda s) \mathrm{d}s = \frac{1}{\Lambda} \big\{1 - \exp(-\Lambda t) \big\}.
\end{split}
\end{align}
The detailed mathematical derivation for (\ref{eq:placebo_components}) is given in the Appendix~\ref{app:competing}.

\subsection{Vaccinated (leaky action) Subjects}
\label{subsec:ve_leaky_components}
Let the vaccine $m$ be a leaky vaccine \citep[page 132]{halloran2010design} with effectiveness $1 - \theta_{i,m}$ against the variant $i$, where $0 \leq \theta_{i,m} \leq 1$. In leaky vaccines, the effectiveness reduces the incidence rate of infection due to variant $i$ in the study period to $\theta_{i,m} \lambda_i$, compared to $\lambda_i$ among placebo. The overall incidence rate of infection by any variant after receiving the vaccine $m$ becomes $\Theta_m \Lambda$ compared to $\Lambda$ in placebo subjects. Here, $1 - \Theta_m = 1 - \sum_{i=1}^I \theta_{i,m} \lambda_i/\Lambda$ is the overall effectiveness of the vaccine $m$ against all variants, with $0 \leq \Theta_m \leq 1$. Under the competing events framework, among subjects vaccinated with the vaccine $m$ ($V=m$), the cumulative-risk of being a control or a case of the variant $i$, and the expected person-time contribution $E(Y\mid V=m)$ is given by:
\begin{align}
\label{eq:leaky_components}
\begin{split}
\mathds{P}(C=0 \mid V=m) &= \exp(-\Theta_m \Lambda t),\\
\mathds{P}(C=i \mid V=m) &= \frac{\theta_{i,m}\lambda_i}{\Theta_m \Lambda} \big\{1 - \exp(-\Theta_m \Lambda t)\big\},\\
E(Y \mid V=m) &= \int_0^t \exp(-\Theta_m \Lambda s) \mathrm{d}s = \frac{1}{\Theta_m \Lambda} \big\{1 - \exp(-\Theta_m \Lambda t) \big\}.
\end{split}
\end{align}
The detailed mathematical derivation for these follows the equations in Appendix~\ref{app:competing}.

\subsection{Vaccinated (all-or-none action) Subjects}
\label{subsec:ve_allornone_components}
To understand our assumed action of all-or-none vaccines \citep[page 132]{halloran2010design} in a multi-variant scenario, consider that there are only three variants $i, j$ and $k$ circulating. Among the subjects vaccinated with the vaccine $m$, let $\theta_{i,m}$ be the proportion of subjects who become immune to the variant $i$ but not to variant $j$ or $k$. Among these subjects infections occur with the combined force of infection of variants $j$ and $k$ given by $\lambda_j + \lambda_k$. We can define $\theta_{j,m}$ and $\theta_{k,m}$ similarly. Next, let $\theta_{ij,m}$ be the proportion of subjects who become immune to both variants $i$ and $j$ but not to variant $k$. Among these subjects infection with variant $k$ happens with the force of infection $\lambda_k$. We can define $\theta_{ik,m}$ and $\theta_{jk,m}$ similarly. Then, let $\theta_{ijk,m}$ be the proportion of subjects who become immune to all three variants. The remaining proportion of subjects are the ones who do not become immune to any variant despite vaccination. For them infections happen with the combined force of infection $\lambda_i + \lambda_j + \lambda_k$ of all three variants. Overall in our three variant example we have seven VE components whose sum is $0\leq (\theta_{i,m} + \theta_{j,m} + \theta_{k,m} + \theta_{ij,m} + \theta_{ik,m} + \theta_{jk,m} + \theta_{ijk,m}) \leq 1$.

For an all-or-none vaccine, as the number of variants increase, the number of VE components $\theta_{\cdot,m}$ increase as well. Consequently, to denote the VE calculations succinctly, we first define $\mathcal{P}(\mathcal{I})$ as the power set of the set $\mathcal{I}$ of all variants circulating. For example, in the scenario above with three variants $i,j,k$, the power set $\mathcal{P}(\mathcal{I}) = \{\{\}, \{i\}, \{j\}, \{k\}, \{i,j\}, \{i,k\}, \{j,k\}, \{i,j,k\}\}$. Then, among the subjects vaccinated with vaccine $m$, let $\theta_{g,m}$ denote the proportion of subjects who are immune to a subset combination $g \in \mathcal{P}(\mathcal{I})$ of variants. Here, when $g=\{\}$, i.e., no variants, then $\theta_{g,m} = \theta_{\{\},m}$ corresponds to the proportion of subjects who are not immune to any variant despite vaccination. On the other hand when $g = \mathcal{I}$, i.e., set of all variants, then $\theta_{g,m} = \theta_{\mathcal{I},m}$ corresponds to the proportion of subjects who are immune to all variants. Subsequently, among subjects vaccinated with the vaccine $m$, the cumulative-risk of being a control or a case of the variant $i$, and the expected person-time contribution $E(Y\mid V=m)$ is given by:
\begin{align}
\label{eq:all_or_none_components}
\begin{split}
\mathds{P}(C=0 \mid V=m) &= \sum_{g \in \mathcal{P}(\mathcal{I})} \mathds{P}(C=0 \mid V=m, G=g) \mathds{P}(G=g)\\
&=\sum_{g \in \mathcal{P}(\mathcal{I})} \exp\bigg\{-\Big(\Lambda - \sum_{l \in g} \lambda_l\Big)t\bigg\} \theta_{g,m}\\
\mathds{P}(C=i \mid V=m) &= \sum_{g \in \mathcal{P}(\mathcal{I} \setminus i)}  \mathds{P}(C=i \mid V=m, G=g) \mathds{P}(G=g)\\
&=\sum_{g \in \mathcal{P}(\mathcal{I} \setminus i)}  \frac{\lambda_i}{\Big(\Lambda - \sum_{l \in g} \lambda_l\Big)} \Bigg[1 - \exp\bigg\{-\Big(\Lambda - \sum_{l \in g} \lambda_l\Big)t\bigg\} \Bigg]\theta_{g,m}\\
E(Y \mid V=m) &= \int_0^t \sum_{g \in \mathcal{P}(\mathcal{I})} \exp\bigg\{-\Big(\Lambda - \sum_{l \in g} \lambda_l\Big)s\bigg\} \theta_{g,m} \mathrm{d}s\\
    &= \sum_{g \in \mathcal{P}(\mathcal{I}) \setminus \mathcal{I}} \frac{\Bigg[1 - \exp\bigg\{-\Big(\Lambda - \sum_{l \in g} \lambda_l\Big)t\bigg\} \Bigg]}{\Big(\Lambda - \sum_{l \in g} \lambda_l\Big)} \theta_{g,m} + t \theta_{I,m},
\end{split}
\end{align}
where, the expansions for $\mathds{P}(C=0 \mid V=m)$ and $\mathds{P}(C=i \mid V=m)$ are obtained using the law of total probability. Specifically, the conditional probability $\mathds{P}(C=0 \mid V=m, G=g)$ denotes the probability of being a control among subjects who are vaccinated with vaccine $m$ and are also immune to variants in the subset combination $g \in \mathcal{P}(\mathcal{I})$. The proportion of these subjects is $\mathds{P}(G=g) = \theta_{g,m}$. Among these subjects the probabilities of being a control or a case of variant $i$ can be derived in the competing events framework following the equations in Appendix~\ref{app:competing}.

The set $\mathcal{P}(\mathcal{I} \setminus i)$ in (\ref{eq:all_or_none_components}) is the power set of the set all variants other than variant $i$. The reason we sum over the power set $\mathcal{P}(\mathcal{I} \setminus i)$ in the probability $\mathds{P}(C=i \mid V=m)$ of being a case of variant $i$, is to exclude those subjects who are immune to variant $i$. Similarly, in the expansion for $E(Y \mid V=m)$, the $\mathcal{P}(\mathcal{I}) \setminus \mathcal{I}$ denotes the power set that excludes the set $\mathcal{I}$ of all variants. To understand these two power sets better, consider the scenario when there are only three variants $i, j$, and $k$. Then $\mathcal{I} = \{i, j, k\}$, and $\mathcal{I} \setminus i = \{j, k\}$. Consequently, $\mathcal{P}(\mathcal{I} \setminus i) = \{\{\}, \{j\}, \{k\}, \{j,k\}\}$ and $\mathcal{P}(\mathcal{I}) \setminus \mathcal{I} = \{\{\}, \{i\}, \{j\}, \{k\}, \{i,j\}, \{i,k\}, \{j,k\}\}$. 

It is important to note that it is not possible to estimate each of the proportion of subjects $\theta_{g,m}$ who are immune to a subset combination $g \in \mathcal{P}(\mathcal{I})$ of variants. This is because, given a total of $I$ variants, the combination $g$ can be any of the $2^I$ combinations from the power set $\mathcal{P}(\mathcal{I})$. However, the total number of equations (\ref{eq:all_or_none_components}) and the corresponding data (Table~\ref{tab:available_data}) are of size $I + 2$. That is, number of equations are less than number of parameters. Although, this is only the case when there are more than two variants ($I > 2$).

%% file: contents/4_ve_time_dependency.tex
\section{Temporal Properties of VE Measures}
\label{sec:ve_time_dependency}
\subsection{Leaky Action Vaccines}
\subsubsection{Variant-specific VE of a Leaky Vaccine $m$ Against Variant $i$}
The variant-specific effectiveness of a leaky vaccine $m$ against variant $i$ is given by $1 - \theta_{i,m}$ (Section~\ref{subsec:ve_leaky_components}). To compare it against the three measures $\mbox{VE}_{i, m}^{irr}, \mbox{VE}_{i, m}^{crr}, \mbox{VE}_{i, m}^{or}$, we combine (\ref{eq:ve_absolute_def}), (\ref{eq:placebo_components}), and (\ref{eq:leaky_components}), to obtain:
\begin{align*}
    \mbox{VE}_{i, m}^{irr} &= 1 - \theta_{i,m},\\
    \mbox{VE}_{i, m}^{crr} &= 1 - \theta_{i,m} \frac{1 - \exp(-\Theta_m \Lambda t)}{\Theta_m \big\{1 - \exp(-\Lambda t)\}},\\
    \mbox{VE}_{i, m}^{or} &= 1 - \theta_{i,m} \frac{\exp(\Theta_m \Lambda t) - 1}{\Theta_m \big\{\exp(\Lambda t) - 1\}}.
\end{align*}
These results show that only the measure $\mbox{VE}_{i, m}^{irr}$ based on incidence rate ratio is time-invariant. This falls in line with the findings of \citet{smith1984assessment}, albeit theirs was a single variant scenario. \citet{gilbert2000comparison} also reached a similar conclusion using a time-to-event model in which VE was expressed in terms of the ratio of cause-specific hazards in vaccinated and placebo groups. 

The measure $\mbox{VE}_{i, m}^{crr}$ based on ratio of cumulative-risks, varies over the study period. In general, $\mbox{VE}_{i, m}^{crr}$ is always smaller than the actual VE $1 - \theta_{i, m}$. This because $\mbox{VE}_{i, m}^{crr}$ can be equivalently rewritten as $(1 - \theta_{i,m}) E(Y \mid V=m)/E(Y \mid V=0)$. Here the ratio $E(Y \mid V=m)/E(Y \mid V=0)$ of person-time is always greater than 1 because, vaccinated subjects should get infected at a slower rate than placebo arm subjects; therefore contributing more person time. When the study length $t$ is very short ($t \to 0)$ and/or the overall incidence rate of infection $\Lambda$ is low ($\Lambda \to 0$), it may happen that their product $\Lambda t \to 0$ as well. In such situation $\mbox{VE}_{i, m}^{crr}$ becomes time-invariant. This is because $\lim_{\Lambda t \to 0}\mbox{VE}_{i, m}^{crr}  = 1 - \theta_{i, m}$. This result is also in line with the findings of \citet{smith1984assessment} in a single variant scenario. The other situation is when $\Lambda t \to \infty$, plausible when the study period is long and/or the pathogen prevalence is high (e.g., an epidemic). In this case $\lim_{\Lambda t \to \infty}\mbox{VE}_{i, m}^{crr} = 1 - \theta_{i, m}/\Theta_m$, depends on the overall effectiveness of the vaccine. 

Lastly, the measure $\mbox{VE}_{i, m}^{or}$ is also time-dependent and always overestimates $1 - \theta_{i, m}$. Although, when $\Lambda t \to 0$ it becomes time-invariant. This is because $\lim_{\Lambda t \to 0}\mbox{VE}_{i, m}^{or} = 1 - \theta_{i, m}$. Conversely, when $\Lambda t \to \infty$ then $\lim_{\Lambda t \to \infty}\mbox{VE}_{i, m}^{or} = 1$. This is an interesting result because when $\Lambda t \to \infty$, then $\mbox{VE}_{i, m}^{or}$ will show a large effectiveness and $\mbox{VE}_{i, m}^{crr}$ may show a small effectiveness.

\subsubsection{Relative VE of a Leaky Vaccine $m$ Against Two Variants $i$ and $j$}
The relative effectiveness of a leaky vaccine $m$ against two variants $i$ and $j$ is given by $1 - \theta_{i,m}/\theta_{j,m}$ (Section~\ref{eq:ve_relative_given_vaccine_def} and Section~\ref{subsec:ve_leaky_components}), where $1 - \theta_{j,m}$ is the effectiveness of vaccine $m$ against variant $j$. To compare it against the three measures $\mbox{VE}_{ij, m}^{irr}, \mbox{VE}_{ij, m}^{crr}, \mbox{VE}_{ij, m}^{or}$, we combine (\ref{eq:ve_relative_given_vaccine_def}), (\ref{eq:placebo_components}), and (\ref{eq:leaky_components}), to obtain:
\begin{equation*}
    \mbox{VE}_{ij, m}^{irr} = \mbox{VE}_{ij, m}^{crr} = \mbox{VE}_{ij, m}^{or} = 1 - \frac{\theta_{i,m}}{\theta_{j,m}}.
\end{equation*}
Thus, all three VE measures are time-invariant. An advantage of this situation is that even if the only data available is counts of infected and not-infected by vaccine and variant, whether via a cohort design or a cumulative case-control design, time-invariant VE can be obtained. Furthermore, another interesting property of these relative VE measures is that even if the incidence rates of infection of the two variants change over the study period, time-invariant VE can be obtained as long as the ratio of the incidence rates of infection of the two variants remains constant (Appendix~\ref{app:varying_hazard_relative_ve}).

\subsubsection{Relative VE of Two Leaky Vaccines $m$ and $n$ Against a Variant $i$}
The relative effectiveness of two vaccines $m$ and $n$ against the same variant $i$ is given by $1 - \theta_{i,m}/\theta_{i,n}$ (Section~\ref{eq:ve_relative_given_vaccine_def} and Section~\ref{subsec:ve_leaky_components}), where $1 - \theta_{i,n}$ is the effectiveness of vaccine $n$ against variant $i$. To compare it against the three definitions $\mbox{VE}_{ij, m}^{irr}, \mbox{VE}_{ij, m}^{crr}, \mbox{VE}_{ij, m}^{or}$ we combine (\ref{eq:ve_relative_given_variant_def}), (\ref{eq:placebo_components}), and (\ref{eq:leaky_components}), to obtain:
\begin{align*}
    \mbox{VE}_{i, mn}^{irr} &= 1 - \frac{\theta_{i,m}}{\theta_{i,n}},\\
    \mbox{VE}_{i, mn}^{crr} &= 1 - \frac{\theta_{i,m}}{\theta_{i,n}} \frac{1 - \exp(-\Theta_m \Lambda t)}{1 - \exp(-\Theta_n \Lambda t)} \frac{\Theta_n}{\Theta_m},\\
    \mbox{VE}_{i, mn}^{or} &= 1 - \frac{\theta_{i,m}}{\theta_{i,n}} \frac{\exp(\Theta_m \Lambda t) - 1}{\exp(\Theta_n \Lambda t) -1} \frac{\Theta_n}{\Theta_m}.
\end{align*}
where, $0 \leq 1 - \Theta_n \leq 1$ is the overall effectiveness of vaccine $n$ against all variants and is defined similar to $\Theta_m$ (Section~\ref{subsec:ve_leaky_components}). The relative VE in this case is time-dependent when either $\mbox{VE}_{i, mn}^{crr}$ or $\mbox{VE}_{i, mn}^{or}$ are used. For both of these, whether the VE increases or decreases over time depends on the relative overall effectiveness $\Theta_n/\Theta_m$ of the two vaccines. A special scenario is when vaccines $m$ and $n$ have different variant-specific effectiveness $\theta_{i,m} \neq \theta_{j,m}$ for variant $i$, but the overall efficacies of the two vaccines are equal $\Theta_m = \Theta_n$. In such a situation, both $\mbox{VE}_{i, mn}^{crr}$ and $\mbox{VE}_{i, mn}^{or}$ are time-invariant and equal to $1 - \theta_{i, m}/\theta_{i, n}$. Furthermore, both $\mbox{VE}_{i, mn}^{crr}$ or $\mbox{VE}_{i, mn}^{or}$ are also time-invariant when $\Lambda t \to 0$ (short study period and/or low infection rate). This is because $\lim_{\Lambda t \to 0}\mbox{VE}_{i, mn}^{crr} = \mbox{VE}_{i, mn}^{or} = 1 - \theta_{i, m}/\theta_{i, n}$. The other scenario is, when the study period is long and/or the infection rate is too high, leading to $\Lambda t \to \infty$. Then, $\lim_{\Lambda t \to \infty}\mbox{VE}_{i, mn}^{crr} = 1 - (\theta_{i, m} \Theta_n) /(\theta_{i, n} \Theta_m)$ and $\lim_{\Lambda t \to \infty}\mbox{VE}_{i, mn}^{or} = 1 - \theta_{i, m} \exp\big\{(\Theta_m - \Theta_n) \infty\big\} /\theta_{i, n}$. Hence, $\mbox{VE}_{i, mn}^{or}$ can be either equal to $1$ if the overall effectiveness of vaccine $m$ is less than that of vaccine $n$, or it can be $-\infty$ if the opposite is true.

%%%%%%%%%%%%%%%%%%%%%%%%%%%%%%%%%%%%%%%%%%%%%%%%%%%%
%% All or none vaccines
%%%%%%%%%%%%%%%%%%%%%%%%%%%%%%%%%%%%%%%%%%%%%%%%%%%%
\subsection{All-or-none Action Vaccines}
\label{subsec:ve_all_so_none}
\subsubsection{Variant-Specific VE of an All-or-none Vaccine $m$ Against Variant $i$}
The variant-specific effectiveness of an all-or-none vaccine $m$ against variant $i$ is the proportion of subjects who are immune to infection from variant $i$, given by $\sum_{g \in \mathcal{P}(\mathcal{I}) \setminus \mathcal{P}(\mathcal{I} \setminus i)} \theta_{g,m}$ (Section~\ref{subsec:ve_allornone_components}). Here, $\mathcal{P}(\mathcal{I}) \setminus \mathcal{P}(\mathcal{I} \setminus i)$ is a power set that only consists of those sets of variants that contain the variant $i$. For example, if there are only three variants $i, j$, and $k$ then $\mathcal{P}(\mathcal{I}) \setminus \mathcal{P}(\mathcal{I} \setminus i) = \{\{i\}, \{i,j\}, \{i,k\}, \{i,j,k\}\}$. The actual corresponding VE for variant $i$ then becomes $\theta_{i,m} + \theta_{ij,m} + \theta_{jk,m} + \theta_{ijk,m}$. For the general scenario with $I$ variants, using (\ref{eq:ve_absolute_def}), (\ref{eq:placebo_components}), and (\ref{eq:all_or_none_components}), the different measures of VE at the end of the study period are given by:
\begin{align*}
    \mbox{VE}_{i,m}^{irr} &= 1 - (1 - \mbox{VE}_{i,m}^{crr}) \times \bigg\{\sum_{g \in \mathcal{P}(\mathcal{I}) \setminus I} \frac{\Lambda \Bigg[1 - \exp\bigg\{-\Big(\Lambda -\sum_{l \in g} \lambda_l\Big)t\bigg\} \Bigg]}{\Big(\Lambda -\sum_{l \in g} \lambda_l\Big) \big\{1 - \exp(-\Lambda t) \big\}} \theta_{g,m} + \frac{\Lambda t \theta_{I,m}}{\big\{1 - \exp(-\Lambda t) \big\}}\bigg\}^{-1},\\
    \mbox{VE}_{i,m}^{crr} &= 1 - \sum_{g \in \mathcal{P}(\mathcal{I} \setminus i)}  \frac{\Lambda \bigg[1 - \exp\Big\{-(\Lambda - \sum_{l \in g} \lambda_l)t\Big\} \bigg]}{\big(\Lambda - \sum_{l \in g} \lambda_l\big) \Big\{1 -  \exp(-\Lambda t)\Big\}}\theta_{g,m},\\
    \mbox{VE}_{i,m}^{or} &= 1 - (1 - \mbox{VE}_{i,m}^{crr}) \times \bigg\{\sum_{g \in \mathcal{P}(\mathcal{I})} \exp\Big(\sum_{l \in g} \lambda_l t\Big) \theta_{g,m}\bigg\}^{-1}.
\end{align*}
These measures of the effectiveness for all-or-none vaccine are time-dependent irrespective of the measure utilized. This is unlike a single variant situation wherein $\mbox{VE}^{crr}$ is known to be time-invariant \citep{smith1984assessment}. In the multi-variant situation $\mbox{VE}_{i,m}^{crr}$ becomes time-invariant only when all $\lambda_k \to 0$, $k\in I/i$. That is, when all variants other than the variant $i$ have a negligible incidence rate of infection. The $\mbox{VE}_{i,m}^{irr}$ is time-invariant only when both $\lambda_k \to 0, k\in I/i$ and the cumulative-incidence rate $\lambda_i t$ of the variant $i$ over the study period $\lambda_i t \to 0$. Along with these conditions if $\lambda_k t \to 0, k\in I/i$ then time-invariant estimates can also be obtained using $\mbox{VE}_{i,m}^{or}$. In general these are very strict conditions.

\subsubsection{Relative VE of an All-or-none Vaccine $m$ Against Two Variants $i$ and $j$}
The relative VE of an all-or-none vaccine $m$ against any two variants $i$ and $j$ can be obtained from Section~\ref{eq:ve_relative_given_vaccine_def} and Section~\ref{subsec:ve_allornone_components} as ${1 -  \{1 - \sum_{g \in \mathcal{P}(\mathcal{I}) \setminus \mathcal{P}(\mathcal{I} \setminus i)} \theta_{g,m}\}/\{1 - \sum_{g \in \mathcal{P}(\mathcal{I}) \setminus \mathcal{P}(\mathcal{I} \setminus j)} \theta_{g,m}\}}$. Here, the $\sum_{g \in \mathcal{P}(\mathcal{I}) \setminus \mathcal{P}(\mathcal{I} \setminus j)} \theta_{g,m}$ corresponds to the variant-specific VE of vaccine $m$ against variant $j$. Using information from (\ref{eq:ve_relative_given_vaccine_def}), we know that our relative VE measures of interest are all equal, i.e., $\mbox{VE}_{ij,m}^{irr} = \mbox{VE}_{ij,m}^{crr} = \mbox{VE}_{ij,m}^{or}$. For brevity they are defined in (\ref{eq:relative_ve_same_vaccine_all_or_none_expansion}) in the Appendix~\ref{app:relative_ve_full_expansions}. The expression in (\ref{eq:relative_ve_same_vaccine_all_or_none_expansion}) is time-dependent. Although, when the incidence rates of infection of all variants are negligible, i.e., $\lambda_i \to 0$, $i\in \mathcal{I}$, then this relative VE becomes time-invariant as well. 

\subsubsection{Relative VE of Two All-or-none Vaccines $m$ and $n$ Against a Variant $i$}
The relative VE of two all-or-none vaccines $m$ and $n$ against the same variant $i$ can be obtained from Section~\ref{eq:ve_relative_given_vaccine_def} and Section~\ref{subsec:ve_allornone_components} as $1 -  \{1 - \sum_{g \in \mathcal{P}(\mathcal{I}) \setminus \mathcal{P}(\mathcal{I} \setminus i)} \theta_{g,m}\}/\{1 - \sum_{g \in \mathcal{P}(\mathcal{I}) \setminus \mathcal{P}(\mathcal{I} \setminus i)} \theta_{g,n}\}$. Here $\sum_{g \in \mathcal{P}(\mathcal{I}) \setminus \mathcal{P}(\mathcal{I} \setminus i)} \theta_{g,n}$ is the effectiveness of vaccine $n$ against variant $i$. To obtain our relative VE measures $\mbox{VE}_{i, mn}^{irr}, \mbox{VE}_{i, mn}^{crr}, \mbox{VE}_{i, mn}^{or}$ we combine (\ref{eq:ve_relative_given_variant_def}), (\ref{eq:placebo_components}), and (\ref{eq:all_or_none_components}). For brevity, the resulting VE expression is given in (\ref{eq:relative_ve_same_variant_all_or_none_expansion}) in Appendix~\ref{app:relative_ve_full_expansions}. Both $\mbox{VE}_{i, mn}^{irr}$ and $\mbox{VE}_{i, mn}^{crr}$ are time-invariant when the incidence rates of infection of all variants are negligible, i.e., $\lambda_i \to 0$, $i\in \mathcal{I}$. In addition if every $\lambda_i t \to 0$, then $\mbox{VE}_{i, mn}^{or}$ becomes time-invariant as well.

%% file: contents/5_test_negative_design.tex
\section{Impact on Test-negative Design (TND) Studies}
\label{sec:tnd}
In this section, we expand upon the results we obtained so far, for the widely used TND study design \citep{lewnard2018measurement}. In TND studies, the aim is to bring in cases and controls through routine surveillance systems. Subjects meeting a clinical case definition are then tested for the disease of interest. For example, in TND studies aiming to estimate effectiveness against influenza, it is common to test and enroll subjects presenting with influenza-like illness \citep{stuurman2020vaccine, rondy20172015}. While cases are subjects infected with the pathogen of interest, controls are subjects who may be infected with other non-target pathogens that elicit similar symptoms. Overall, the premise of the TND studies is that by sampling subjects who seek medical care, healthcare-seeking behavior bias can be limited. Depending on the exact schedule to sample controls, TND design imitates case-control design with incidence density sampling \citep{dean2019re} or inclusive sampling \citep{vandenbroucke2012case}. 

Consider that the size of our population of interest is $n$, wherein an infection can occur either from a variant $i \in \mathcal{I}$ of the pathogen of interest ($C=i$) or from a non-target pathogen ($C=\Omega)$. We assume that infection with the target pathogen or one of the non-target pathogens does not lead to immunity against other non-target pathogens \citep{dean2020temporal}. Following \citet{lewnard2018measurement} we consider that an infected subject may be recorded as a case or as control only if symptoms appear and the subject subsequently seeks care and gets tested. We denote symptom appearance by $S=1$ (or 0) and seeking care and getting tested by $Z=1$ (or 0). Suppose, after getting tested, the subject is found to be infected with a variant $i \in \mathcal{I}$ of target pathogen. In that case, the subject is counted as a case. In contrast, if the subject is infected with a non-target pathogen, the subject is counted as a control. We next present equations for the expected number of variant-specific cases and expected number of controls. These expressions will be later employed in the expression for VE in TND studies.

\subsection{Expected Number of Variant-specific Cases in TND Studies}
We denote the expected number of cases of variant $i$ of the pathogen of interest who were vaccinated with vaccine $m$ by $E(N_{i,m})$, and the ones who were unvaccinated by $E(N_{i,0})$. Following the work of \citet{lewnard2018measurement} we define the cases as,
\begin{align}
\label{eq:tnd_cases}
\begin{split}
E(N_{i,m}) &=n \mathds{P}(Z=1, S=1, C=i, V=m)\\
&=n \mathds{P}(Z=1 \mid S=1, V=m) \mathds{P}(S=1 \mid C=i) \mathds{P}(C=i \mid V=m) \mathds{P}(V=m),\\
E(N_{i,0}) &=n \mathds{P}(Z=1, S=1, C=i, V=0)\\
&=n \mathds{P}(Z=1 \mid S=1, V=0) \mathds{P}(S=1 \mid C=i) \mathds{P}(C=i \mid V=0) \mathds{P}(V=0).
\end{split}
\end{align}
Here, $\mathds{P}(Z=1, S=1, C=i, V=m)$ is the joint probability that over the study period, a subject seeks care ($Z=1$) after having symptoms ($S=1$) due to infection with variant $i$, and received vaccine $m$ before the study began. The probability $\mathds{P}(Z=1, S=1, C=i, V=0)$ has a similar interpretation and is applicable for unvaccinated subjects ($V=0$). 

In (\ref{eq:tnd_cases}), we have split the joint probability $\mathds{P}(Z=1, S=1, C=i, V=m)$ into conditional probabilities, and especially into $\mathds{P}(Z=1 \mid S=1, V=m)$ and $\mathds{P}(S=1 \mid C=i)$. The importance of $\mathds{P}(Z=1 \mid S=1, V=m)$ is that it means that seeking care ($Z=1$) is conditionally independent of the type of infection given the vaccination status and symptoms. That is, $\mathds{P}(Z=1 \mid S=1, V=m) = \mathds{P}(Z=1 \mid S=1, C=i, V=m)$. This also reflects what happens in practice in a TND study. Specifically, a subject may base their decision to seek care knowing their vaccination status and symptoms even though they may not know their infection type until getting tested. The probability $\mathds{P}(S=1 \mid C=i)$ indicates that the occurrence of symptoms depends on the type of infection and is conditionally independent of the vaccination status. As we show later in this section, these conditional probabilities cancel out in VE calculations.

\subsection{Expected Number of Controls in TND Studies}
Consider that the instantaneous rate of being infected with any non-target pathogen ($C=\Omega$) is a constant $\lambda_\Omega$. This rate $\lambda_\Omega$ does not depend upon getting vaccinated with vaccine $m$ for the target pathogen. Consequently, even if a subject becomes a case of a variant $i$ of the pathogen of interest, the subject can also later become a case of non-target pathogens multiple times over the study period. Whether such a subsequent infection with the non-target pathogen qualifies to be counted as a control, depends upon the sampling design. In this regard, two designs of interest for us are inclusive sampling and incidence density sampling. We present them next.

\subsubsection{Inclusive Sampling}
In inclusive sampling a subject is counted as control as many times as they get infected with the non-target pathogen, seek care and get tested. Then, over the study period $[0, t]$ the expected number of controls $E(N_{\Omega,m})$ vaccinated with vaccine $m$ and expected number of unvaccinated controls $E(N_{\Omega,0})$ are given by \citep{lewnard2018measurement}:
\begin{align}
\label{eq:tnd_inclusive_controls}
\begin{split}
E(N_{\Omega,m}) &= n \int_0^t \mathds{P}(Z=1, S=1, V=m \mid C=\Omega) \lambda_\Omega \mathrm{d}u\\
&= n t \mathds{P}(Z=1 \mid S=1, V=m) \mathds{P}(S=1 \mid C=\Omega) \mathds{P}(V=m) \lambda_\Omega,\\
E(N_{\Omega,0}) &= n \int_0^t \mathds{P}(Z=1, S=1, V=0 \mid C=\Omega) \lambda_\Omega \mathrm{d}u\\
&= nt \mathds{P}(Z=1 \mid S=1, V=0) \mathds{P}(S=1 \mid C=\Omega) \mathds{P}(V=0) \lambda_\Omega.
\end{split}
\end{align}
Here, $\mathds{P}(Z=1, S=1, V=m \mid C=\Omega)$ is the probability that a subject known to be infected with a non-target pathogen seeks care after having symptoms and received vaccine $m$ before the study began. The interpretation of the probability $\mathds{P}(Z=1, S=1, V=0 \mid C=\Omega)$ is similar, but it is applicable to unvaccinated subjects.

\subsubsection{Incidence Density Sampling}
In incidence density sampling, after a subject becomes a case of a variant $i$ of the pathogen of interest ($C=i$), they leave the risk set. Thus any subsequent infection with a non-target pathogen is ignored. Although, before this censoring time, the subject can be counted as control as many times as they get infected with the non-target pathogen, seek care, and get tested. We have previously defined the expected time of getting infected with the pathogen of interest as $E(Y \mid V=m)$ and $E(Y \mid V=0)$ among subjects vaccinated with vaccine $m$ and unvaccinated subjects, respectively. By replacing $t$ with with $E(Y \mid V=\cdot)$ in (\ref{eq:tnd_inclusive_controls}), we obtain $E(N_{\Omega,m})$ and  $E(N_{\Omega,0})$ in an incidence sampling approach. They are given by \citep{dean2019re}:
\begin{align}
\label{eq:tnd_incidence_sampling_controls}
\begin{split}
E(N_{\Omega,m}) &=n E(Y \mid V=m) \mathds{P}(Z=1 \mid S=1, V=m) \mathds{P}(S=1 \mid C=\Omega) \mathds{P}(V=m) \lambda_\Omega,\\
E(N_{\Omega,0}) &= n E(Y \mid V=0) \mathds{P}(Z=1 \mid S=1, V=0) \mathds{P}(S=1 \mid C=\Omega) \mathds{P}(V=0) \lambda_\Omega.
\end{split}
\end{align}
A problem in the incidence density sampling is that ascertaining if a subject is indeed a control, i.e., they have not had a previous infection with the target pathogen, can be challenging \citep{reply_dean2019}.

\subsection{Vaccine Effectiveness in TND Studies}
\subsubsection{Variant-specific VE of a Vaccine $m$ Against Variant $i$}
Let $\mbox{VE}_{i,m}^{TND}$ be the variant-specific effectiveness of vaccine $m$ against variant $i$ of the pathogen of interest in a TND study. It is defined in terms of ratio of odds of being vaccinated with vaccine $m$ among the subjects infected with variant $i$, and the same odds in subjects infected with non-target pathogen. Specifically, 
\begin{equation*}
\mbox{VE}_{i,m}^{TND} = 1 - \frac{E(N_{i,m})}{E(N_{i,0})} \div \frac{E(N_{\Omega,m})}{E(N_{\Omega,0})}
\end{equation*}
This expression can be further expanded for both inclusive and incidence density sampling designs using (\ref{eq:tnd_inclusive_controls}) and (\ref{eq:tnd_incidence_sampling_controls}), respectively. We then obtain,
\begin{align}
\label{eq:tnd_absolue_ve}
    \begin{split}
        \mbox{VE}_{i,m}^{TND} &= 1 - \frac{\mathds{P}(C=i \mid V=m)}{\mathds{P}(C=i \mid V=0)} = \mbox{VE}_{i,m}^{crr}, \text{in inclusive sampling},\\
        \mbox{VE}_{i,m}^{TND} & = 1 - \frac{\mathds{P}(C=i \mid V=m)}{E(Y \mid V=m)} \div \frac{\mathds{P}(C=i \mid V=0)}{E(Y \mid V=0)} = \mbox{VE}_{i,m}^{irr}, \text{in incidence density sampling}.
    \end{split}
\end{align}
Thus, the temporal properties of the VE measures $\mbox{VE}_{i,m}^{crr}$ and $\mbox{VE}_{i,m}^{irr}$ described in Section~\ref{sec:ve_time_dependency} also hold in TND studies. Among these, a peculiar one is that when a vaccine has an all-or-none action, then irrespective of the sampling design $\mbox{VE}_{i,m}^{TND}$ will always be time-dependent (Section~\ref{subsec:ve_all_so_none}). This is in contrast to the findings to \citet{lewnard2018measurement} in TND studies with a single variant. The only scenario in which $\mbox{VE}_{i,m}^{TND}$ is time-invariant is when the vaccine has a leaky action and an incidence density sampling design is used. 

\subsubsection{Relative VE of a Vaccine $m$ Against Two Variants $i$ and $j$}
Let the relative VE of a vaccine $m$ against two variants $i$ and $j$ be denoted by $\mbox{VE}_{ij,m}^{TND}$. Using (\ref{eq:tnd_absolue_ve}) and the relative VE definitions from Section~\ref{sec:ve_measures} we can show that $\mbox{VE}_{ij,m}^{TND} = \mbox{VE}_{ij,m}^{crr}$ for inclusive sampling and $\mbox{VE}_{ij,m}^{TND} = \mbox{VE}_{ij,m}^{irr}$ for incidence density sampling. In Section~\ref{sec:ve_time_dependency} we have shown that both $\mbox{VE}_{ij,m}^{crr}$ and $\mbox{VE}_{ij,m}^{irr}$ are time-invariant for leaky vaccines and time-dependent for all-or-none vaccines. Consequently, irrespective of the sampling design $\mbox{VE}_{ij,m}^{TND}$ should be time-invariant for leaky vaccines but not for all-or-none vaccines. 

\subsubsection{Relative VE of Two Vaccines $m$ and $n$ Against a Variant $i$}
Let the relative VE of two vaccines $m$ and $n$ against a variant $i$ be $\mbox{VE}_{i,mn}^{TND}$. Using (\ref{eq:tnd_absolue_ve}) and the relative VE definitions from Section~\ref{sec:ve_measures} we can show that $\mbox{VE}_{i,mn}^{TND} = \mbox{VE}_{i,mn}^{crr}$ for inclusive sampling and $\mbox{VE}_{i,mn}^{TND} = \mbox{VE}_{i,mn}^{irr}$ for incidence density sampling. Based on the properties of $\mbox{VE}_{i,mn}^{crr}$ and $\mbox{VE}_{i,mn}^{irr}$ we can say that $\mbox{VE}_{i,mn}^{TND}$ is time-invariant only in leaky vaccines, that too only if the incidence density sampling is used.

%% file: contents/6_sample_size.tex
\section{Sample Size Calculation}
\label{sec:samp_size}
An important problem in effectiveness studies is sample size calculation during the study planning phase. In this regard, investigators may face two types of challenges. First, to have enough subjects to show a non-zero VE, wherein one can obtain an appropriate sample size utilizing the hypothesis testing framework. The second type of challenge is having enough subjects to obtain precise estimates of VE to further assist in decisions pertaining to the various phases of the development of a vaccine. In statistical terms, precision \citep{kelley2003obtaining} refers to confidence limits of VE. In general, narrower confidence intervals can be obtained with larger sample sizes. While the actual confidence limits depend upon the actual data, using Monte Carlo simulations, confidence intervals can also be simulated, and a range for both the upper and lower confidence limit can be obtained. To resolve the aforementioned challenges and to meet the requirements of the DRIVE project (Section~\ref{sec:intro}), we developed a sample size calculator. It is available at \url{https://apps.p-95.com/drivesamplesize/}, and the underlying calculations are also implemented in an \textbf{R} package available at \url{https://github.com/anirudhtomer/vess}. It is important to note that for VE based on incidence rate ratio, currently our \textbf{R} package only supports leaky-action vaccines.

\subsection{Minimum Detectable VE}
When the aim is to find a minimum detectable VE for a given sample size, we utilize the framework of hypothesis testing. The null hypothesis we utilize is that the $\mbox{VE} = 0$ and the alternate hypothesis is that the $\mbox{VE} \neq 0$. In cohort studies with a given sample size $n$, when the interest is in variant-specific or relative VE measures based on ratio of cumulative-risks, then we utilize the methodology proposed by \citet{woodward2013epidemiology} to find the minimum detectable VE. Whereas, when the interest is in variant-specific or relative VE measures based on incidence rate ratio then we use the methodology proposed by \citet{lwanga1991sample}. Both methodologies require vaccine coverage and the number of study subjects. However, these methodologies are only suitable for a single vaccine and single variant scenario. To use them in our multi-variant and multi-vaccine scenario, we recalculate vaccine coverage in terms of only the subjects that are used in a particular calculation. For example, for variant-specific VE of vaccine $m$ against variant $i$, we define coverage as $\{\mathds{P}(C=i, V = m) + \mathds{P}(C=0, V = m)\} / \{\mathds{P}(C=i, V = m) + \mathds{P}(C=0, V = m) + \mathds{P}(C=i, V = 0) + \mathds{P}(C=0, V = 0)\}$. The total subjects are calculated as $n \{\mathds{P}(C=i, V = m) + \mathds{P}(C=0, V = m) + \mathds{P}(C=i, V = 0) + \mathds{P}(C=0, V = 0)\}$. That is, we ignore all subjects that are vaccinated with vaccines other than vaccine $m$ or are infected with any variant other than variant $i$. Similarly, for relative VE of two vaccines $m$ and $n$ against a variant $i$, we ignore all subjects that are unvaccinated or are vaccinated with vaccines other than vaccine $m$, or are infected with any variant other than variant $i$.

For a classic case-control study and TND study (with inclusive sampling), let the total cases be $x$, and the controls per case be $r$. Thus total controls are given by $rx$. The measures of interest are those based on ratio of cumulative-risks, namely variant-specific $\mbox{VE}^{crr}_{i,m}$ or the relative $\mbox{VE}^{crr}_{i, mn}$. These measures are estimated using the odds ratio (Section~\ref{sec:tnd}). Here, to calculate the minimum detectable VE we employ the methodology of \citet{dupont1988power}. However, this methodology is only suitable for a single vaccine and single variant scenario. To use it in our multi-variant and multi-vaccine scenario, we recalculate controls per case in terms of only the subjects that are used in a particular calculation. For example, for variant-specific VE of vaccine $m$ against variant $i$, we first define total cases for the variant-specific VE as $x \{\mathds{P}(V=m, C=i \mid C\neq 0)+\mathds{P}(V=0, C=i \mid C\neq 0)\}$. Here conditioning on $C\neq 0$ indicates that these probabilities are relevant only for cases. The total controls are then given by $x\times r \{\mathds{P}(V=m \mid C=0) + \mathds{P}(V=0 \mid C=0)\}$. Subsequently, we recalculate the controls per case using the aforementioned new totals for controls and cases.

\subsection{Expected Precision of VE}
When the aim is to find the expected precision of VE, we use a simulation-based approach. Specifically, in cohort studies, in each simulation round we first sample the data $n_{i,m}$ of the number of subjects infected with variant $i$ and vaccinated with vaccine $m$. This data is sampled using a multinomial distribution $\mathds{P}(C=i, V=m)$ of the probabilities of being infected by variant $i$ (or being uninfected $C=0$) and being vaccinated with the vaccine $m$ (or placebo $V=0$). Subsequently, the overall sample size $n$ is given by $n = \sum_{i=0}^{I}\sum_{m=0}^{M}n_{i,m}$. 

When the VE measures based on incidence rate ratio, namely variant-specific $\mbox{VE}_{i,m}^{irr}$ and the relative $\mbox{VE}_{i,mn}^{irr}$ are of interest, then in each simulation round we also sample the person-time contributions $Y \mid V=m$ for each vaccine and $Y \mid V=0$ for placebo subjects from an exponential distribution. Among the placebo subjects, the rate parameters of this exponential distribution is simply the incidence rate of infection $\lambda_i$. Whereas, in subjects vaccinated with vaccine $m$, the corresponding rate parameter is $\theta_{i,m} \lambda_i$. In this regard, both $\theta_{i,m}$ and $\lambda_i$ are needed as input parameters. To subsequently estimate variant-specific and relative VE we use the sampled data $n_{i,m}$ and $Y \mid V=m$ to estimate incidence rate ratios. The standard errors and confidence intervals are estimated using normal approximation to the distribution of log incidence rate ratio \citet{halloran2010design}. For VE measures based on the ratio of cumulative-risks
we only use the sampled data $n_{i,m}$, and for estimating the standard error and confidence intervals we use the methodology of \citet{morris1988statistics}. In each simulation round, we adjust standard errors for potential confounders using the approach of \citet{hsieh2000sample}. The confidence intervals obtained over multiple such simulations are then averaged to find the expected upper and expected lower limits of the confidence interval. The simulation approach to calculate precision is not currently used by existing calculators, which instead provide the confidence interval from a single cross table of infection and vaccination status, with cells values $E\{n_{i,m}\}=n \mathds{P}(C=i, V=m)$. Such a confidence interval does not represent the expected precision but rather represents precision for expected cell counts, thus ignoring the randomness of the data generating process.

For case-control and TND studies (with inclusive sampling), the total number of cases of $x$ and the controls per case $r$ are required as input parameters. In each simulation round we then sample total cases as $x = \sum_{i=i}^{I}\sum_{m=0}^{M}x_{i,m}$. Here, $x_{i,m}$ are the cases sampled from the multinomial distribution $\mathds{P}(V=m, C=i\mid C\neq 0)$ of the probabilities of being vaccinated by vaccine $m$ and being a case of variant $i$ among the cases. Similarly, a total of $rx$ controls are also sampled from the multinomial distribution $\mathds{P}(V=m \mid C=0)$. Subsequently, we utilize the sample odds ratio as the estimator of VE (Section~\ref{sec:tnd}). The standard error and confidence interval are obtained using the approach of \citet{morris1988statistics}. The rest of the simulation approach remains similar to the cohort studies. 

\subsection{Temporal Properties of VE in an \textbf{R} Package}
In Section~\ref{sec:ve_time_dependency} we observed that various VE measures can be time-invariant or time-dependent. However, the degree of time-dependency varies as per vaccine action, the length of the study period $t$, and the force of infection of variants $\lambda_i$. The temporal properties of relative VE  may also depend on the relative overall effectiveness of brands $\Theta_m/\Theta_n$. To assist practitioners in quantitatively assessing the amount of time-dependency they should expect, we implemented the variant-specific and relative VE definitions for both leaky and all-or-none vaccines in our \textbf{R} package available at \url{https://github.com/anirudhtomer/vess}. 

%% file: contents/7_discussion.tex
\section{Discussion}
\label{sec:discussion}
In this paper, we worked on defining and evaluating the temporal properties of measures of the effectiveness of vaccines in the presence of multiple pathogen variants and multiple vaccines. We focused on cohort and test-negative design (TND) studies. The reason why the multi-variant and multi-vaccine scenario faces unique challenges is because of the competing-events/variants \citep{putter2007tutorial} situation. We worked on two challenges. First, defining variant-specific VE measures and measures of relative VE of vaccines, and evaluating the conditions and the extent to which a VE measure can be time-varying. Second, conducting sample size calculations in a multi-variant and multi-vaccine scenario. We concentrated on VE measures based on incidence rate ratio ($\mbox{VE}^{irr}$), cumulative-risk ratio ($\mbox{VE}^{crr}$), and odds ratio ($\mbox{VE}^{or}$). 

The issue that some measures of VE can be time-varying over the study period depending on the vaccine's mode of action has been discussed by \citet{smith1984assessment}, albeit only in a single variant and single vaccine scenario. In the presence of multiple variants and vaccines, the variant-specific and relative VE of vaccines can both be time-dependent. In this regard, we show that when a vaccine has an all-or-none action, then all measures ($\mbox{VE}^{irr}$, $\mbox{VE}^{crr}$, $\mbox{VE}^{or}$) of variant-specific VE of a vaccine are time-dependent. This is in contrast to the findings of \citet{smith1984assessment} and \citet{lewnard2018measurement} who showed that $\mbox{VE}^{crr}$ is time-invariant for all-or-none vaccines when there is only one variant and one vaccine. Also, for all-or-none vaccines, every measure of relative VE of a vaccine against any two variants, and every measure of relative VE of two vaccines against a particular variant, is time-dependent. However, our findings for leaky vaccines are similar to those of \citet{smith1984assessment}. Specifically, VE measures based on incidence ratio are time-invariant for leaky vaccines. In addition, all measures of relative VE a vaccine against any two variants is also time-invariant. We evaluated the temporal properties effectiveness of vaccines for both cohort and TND studies. For TND studies, extending upon the work on \cite{lewnard2018measurement} and \citet{dean2019re} for variant-specific VE and relative VE, we showed that odds ratio from TND studies estimate $\mbox{VE}^{crr}$ and $\mbox{VE}^{irr}$ when inclusive and incidence density sampling is used, respectively. Thus the temporal properties of $\mbox{VE}^{crr}$ and $\mbox{VE}^{irr}$ also hold in TND studies.

Our findings are relevant for practitioners. First, practitioners will benefit from a smaller study period for vaccines known to have an all-or-none action. This is because, for all-or-none vaccines, the variant-specific VE measures can be time-invariant when the study period is short. In general, for both leaky and all-or-none vaccines, when the disease incidence rate is low, a larger cohort of subjects followed up for a shorter period may provide time-invariant estimates than a smaller cohort followed up for a more extended period. In practice, VE may vary due to the varying incidence rate of infection, waning VE, time-dependence of the chosen VE measure. Consequently, discerning the cause can be difficult when VE varies over time. In this regard, the time-dependence of a VE measure also depends on the vaccine action. Practitioners may utilize the measure for the relative VE of a vaccine against any two variants to identify the vaccine action. This is because this relative VE remains time-invariant in leaky vaccines but not in all-or-none action vaccines. Besides, the relative VE of a vaccine against two variants is time-invariant for leaky vaccines even when the incidence rate of infection changes over time.

To assist practitioners in evaluating the extent to which a chosen measure of VE can be time-dependent and to help them in finding an appropriate sample size, we created an \textbf{R} package. The package allows sample size calculations for two situations: hypothesis testing and aiming for a certain precision of the VE estimate. Our package has four key features not available in current calculators. First, we support the multi-variant and multi-vaccine scenario, which is very relevant in Influenza and the current COVID-19 pandemic. The package supports both the cohort design and TND studies. Second, along with sample size calculations for variant-specific VE, we also implemented calculations for relative VE. Third, we adjust the sample size calculations for potential confounders using the methodology proposed by \citet{hsieh2000sample}. Last, for the precision of VE, we utilize simulations rather than the expected value of cell counts.

A limitation of our work is that we assumed constant incidence rates of infection, which may be difficult to justify in practice, and especially over more extended study periods. Although, our motivation to choose constant rates of infection was to show that some measures of VE vary over time even when the incidence rates of infection are constant. Our work does not have any data-based illustrations. Although, using our \textbf{R} package, practitioners may evaluate the extent to which their selected VE measures are time-varying. We also did not focus on VE measures based on hazard ratio. Given the competing-event situation with multiple variants, discussing competing risk models and cause-specific hazard ratio as a measure of VE would have been interesting. We also assumed that infection with a particular variant inhibits infection from other pathogen variants over the study period. Working under a general framework wherein reinfections are allowed will be a natural extension of this work. 

%% file: contents/8_acknowledgments.tex
\section*{Acknowledgements}
We would like to express our gratitude to Paddy Farrington for sharing ideas during the initial development of the manuscript, subsequent review, and discussing limitations of this work. We are grateful to Jos Nauta from Abbott Laboratories for reviewing our manuscript and recommending important changes in some of terminologies that we used. Lastly, we would like to thank Kaatje Bollaerts and Anke Stuurman from P95 for coordinating and discussing the epidemiological aspects of this project. 

%% file: contents/appendix.tex
\section{Cumulative-risk and Person-time Calculations In a Competing-variants Scenario}
\label{app:competing}
Suppose that given a certain vaccination status $V$ a subject a prone to infection by a set of variants $\mathcal{I} = \{1, \ldots, I\}$. The combined force of infection by all variants is given by $\Lambda = \sum_{i\in \mathcal{I}} \lambda_i$. In this situation the subjects who do not get infected by any variant over the study period $[0, t]$ are control subjects. For these subjects the probability of being a control is same as in \citet[Equation~11]{putter2007tutorial}. It is given by,
\begin{equation*}
    \mathds{P}(C=0 \mid V) = \exp(-\int_0^t \sum_{i\in \mathcal{I}} \lambda_i \mathrm{d}s) = \exp(-\Lambda t)
\end{equation*}
Following standard survival analysis methods, the expected person-time contribution of a subject $E(Y \mid V)$ is equal to the restricted mean survival time \citep{zhao2016restricted}. This restricted mean survival time is defined as the area under the probability of being a control over the study-period $[0, t]$ \citep{zhao2016restricted}. Thus, $E(Y \mid V)$ can be defined as, 
\begin{equation*}
    E(Y \mid V) = \int_0^t \mathds{P}(C=0 \mid V) \mathrm{d}s =  \int_0^t \exp(-\Lambda s) \mathrm{d}s = \frac{1}{\Lambda} \big\{1 - \exp(-\Lambda t) \big\}.
\end{equation*}
Lastly, the probability of being a case of a certain variant $i$ is same as in \citet[Equation~12]{putter2007tutorial}. It is given by,
\begin{equation*}
    \mathds{P}(C=i \mid V) = \int_0^t \lambda_i \mathds{P}(C=0 \mid V) \mathrm{d}s = \lambda_i \int_0^t  \exp(-\Lambda s) \mathrm{d}s = \frac{\lambda_i}{\Lambda} \big\{1 - \exp(-\Lambda t) \big\}
\end{equation*}

\section{Relative VE of a Leaky Vaccine $m$ Against Two Variants $i$ and $j$ When the Force of Infection of the Variants is Varying}
\label{app:varying_hazard_relative_ve}
Suppose that the force of infection of a variants $i$ and $j$ varies over time and is given by $\lambda_i (s)$ and $\lambda_i (s)$ at time $s$, respectively. Let ratio of forces of infection of the two variants be a constant $f = \lambda_i(s)/\lambda_j(s)$ over the study period. Let there be a leaky vaccine $m$ that reduces the force of infection of variant $i$ to $\theta_{i,m}\lambda_i (s)$ and of variant $j$ to $\theta_{j,m}\lambda_j (s)$. Then the relative VE of vaccine $m$ against two variants $i$ and $j$ after follow-up over the period $[0, t]$ remains time-invariant. Furthermore, it can be estimated from data that is only counts of infected and not-infected by vaccine and variant. The data can be collected based on either a cohort design or a cumulative case-control design. The proof of this result is given by extending (\ref{eq:ve_relative_given_vaccine_def}) as,
\begin{align}
\begin{split}
    \mbox{VE}_{ij,m}^{irr} = \mbox{VE}_{ij,m}^{crr} = 
    \mbox{VE}_{ij,m}^{or} &= 1 - \frac{\mathds{P}(C=i \mid V=m)}{\mathds{P}(C=j \mid V=m)} \div \frac{\mathds{P}(C=i \mid V=0)}{\mathds{P}(C=j \mid V=0)}\\
    & = 1 - \frac{\int_0^t \theta_{i,m}\lambda_i (s) S_m(s) \mathrm{d}s}{\int_0^t \theta_{j,m}\lambda_j (s) S_m(s) \mathrm{d}s} \div \frac{\int_0^t \lambda_i (s) S_0(s) \mathrm{d}s}{\int_0^t \lambda_j (s) S_0(s)\mathrm{d}s}\\
    & = 1 - \frac{\int_0^t \theta_{i,m} f \lambda_j (s) S_m(s) \mathrm{d}s}{\int_0^t \theta_{j,m}\lambda_j (s) S_m(s) \mathrm{d}s} \div \frac{\int_0^t f \lambda_j (s) S_0(s) \mathrm{d}s}{\int_0^t \lambda_j (s) S_0(s)\mathrm{d}s}\\
    & = 1 - \frac{\theta_{i,m}}{\theta_{j,m}} \frac{f\int_0^t  \lambda_j (s) S_m(s) \mathrm{d}s}{\int_0^t \lambda_j (s) S_m(s) \mathrm{d}s} \div \frac{f\int_0^t \lambda_j (s) S_0(s) \mathrm{d}s}{\int_0^t \lambda_j (s) S_0(s)\mathrm{d}s}\\
    & = 1 - \frac{\theta_{i,m}}{\theta_{j,m}}
\end{split}
\end{align}
where $S_m(s) = \mathds{P}(C=0 \mid V=m)$ is the probability of being a control at time $s$ among subjects vaccinated with vaccine $m$ and $S_0(s) = \mathds{P}(C=0 \mid V=0)$ is the probability of being a control at time $s$ among placebo subjects. 

\section{Full Expansions of Relative VE for All-or-none Action Vaccines}
\label{app:relative_ve_full_expansions}
Using (\ref{eq:placebo_components}), and (\ref{eq:all_or_none_components}), the relative VE of the vaccine $m$ against any two variants $i$ and $j$ is obtained as,
\begin{align}
\label{eq:relative_ve_same_vaccine_all_or_none_expansion}
\begin{split}
    \mbox{VE}_{ij,m}^{irr} = \mbox{VE}_{ij,m}^{crr} = \mbox{VE}_{ij,m}^{or} = 1 &- \Bigg\{\sum_{g \in \mathcal{P}(\mathcal{I} \setminus i)}  \frac{\Lambda \bigg[1 - \exp\Big\{-(\Lambda - \sum_{l \in g} \lambda_l)t\Big\} \bigg]}{\big(\Lambda - \sum_{l \in g} \lambda_l\big) \Big\{1 -  \exp(-\Lambda t)\Big\}}\theta_{g,m}\Bigg\}\\&\quad \times \Bigg\{\sum_{g \in \mathcal{P}(\mathcal{I} \setminus j)}  \frac{\Lambda \bigg[1 - \exp\Big\{-(\Lambda - \sum_{l \in g} \lambda_l)t\Big\} \bigg]}{\big(\Lambda - \sum_{l \in g} \lambda_l\big) \Big\{1 -  \exp(-\Lambda t)\Big\}}\theta_{g,m}\Bigg\}^{-1}
\end{split}
\end{align}

Using (\ref{eq:ve_relative_given_variant_def}), (\ref{eq:placebo_components}), and (\ref{eq:all_or_none_components}) the relative VE of two vaccines $m$ and $n$ against the same variant $i$ is given by,
\begin{align}
\label{eq:relative_ve_same_variant_all_or_none_expansion}
\begin{split}
\mbox{VE}_{i, mn}^{irr} &= 1 - (1-\mbox{VE}_{i, mn}^{crr}) \bigg\{\sum_{g \in \mathcal{P}(\mathcal{I}) \setminus I} \frac{\Lambda \Bigg[1 - \exp\bigg\{-\Big(\Lambda -\sum_{l \in g} \lambda_l\Big)t\bigg\} \Bigg]}{\Big(\Lambda -\sum_{l \in g} \lambda_l\Big) \big\{1 - \exp(-\Lambda t) \big\}} \theta_{g,n} + \frac{\Lambda t \theta_{I,n}}{\big\{1 - \exp(-\Lambda t) \big\}}\bigg\}\\& \quad \quad \quad \times \bigg\{\sum_{g \in \mathcal{P}(\mathcal{I}) \setminus I} \frac{\Lambda \Bigg[1 - \exp\bigg\{-\Big(\Lambda -\sum_{l \in g} \lambda_l\Big)t\bigg\} \Bigg]}{\Big(\Lambda -\sum_{l \in g} \lambda_l\Big) \big\{1 - \exp(-\Lambda t) \big\}} \theta_{g,m} + \frac{\Lambda t \theta_{I,m}}{\big\{1 - \exp(-\Lambda t) \big\}}\bigg\}^{-1}\\
\mbox{VE}_{i, mn}^{crr} &= 1 - \Bigg\{\sum_{g \in \mathcal{P}(\mathcal{I} \setminus i)}  \frac{\Lambda \bigg[1 - \exp\Big\{-(\Lambda - \sum_{l \in g} \lambda_l)t\Big\} \bigg]}{\big(\Lambda - \sum_{l \in g} \lambda_l\big) \Big\{1 -  \exp(-\Lambda t)\Big\}}\theta_{g,m}\Bigg\}\\ & \quad \quad \quad \times \Bigg\{\sum_{g \in \mathcal{P}(\mathcal{I} \setminus i)}  \frac{\Lambda \bigg[1 - \exp\Big\{-(\Lambda - \sum_{l \in g} \lambda_l)t\Big\} \bigg]}{\big(\Lambda - \sum_{l \in g} \lambda_l\big) \Big\{1 -  \exp(-\Lambda t)\Big\}}\theta_{g,n}\Bigg\}^{-1}\\
\mbox{VE}_{i, mn}^{or} &= 1 - (1-\mbox{VE}_{i, mn}^{crr}) \bigg\{\sum_{g \in \mathcal{P}(\mathcal{I})} \exp\Big(\sum_{l \in g} \lambda_l t\Big) \theta_{g,n}\bigg\} \bigg\{\sum_{g \in \mathcal{P}(\mathcal{I})} \exp\Big(\sum_{l \in g} \lambda_l t\Big) \theta_{g,m}\bigg\}^{-1}
\end{split}
\end{align}